# Deep-learning real-time phase retrieval of imperfect diffraction patterns from X-ray free-electron lasers

Sung Yun Lee,[1,2,3] Do Hyung Cho,[1,2] Chulho Jung,[1,2] Daeho Sung,[1,2] Daewoong Nam,[4] Sangsoo Kim,[4] Changyong Song[1,2,3,*]

[1]*Department of Physics, POSTECH, Pohang 37673, Republic of Korea,* [2]*Photon Science Center, POSTECH, Pohang 37673, Republic of Korea,* [3]*Center for Ultrafast Science on Quantum Matter, Max Planck POSTECH/Korea Research Initiative, Pohang 37673, Republic of Korea,* [4]*Pohang Accelerator Laboratory, POSTECH, Pohang 37673, Republic of Korea*

\* *Corresponding author; email: cysong@postech.ac.kr*

**Abstract**

Machine learning is attracting surging interest across nearly all scientific areas by enabling the analysis of large datasets and the extraction of scientific information from incomplete data. Data-driven science is rapidly growing, especially in X-ray methodologies, where advanced light sources and detection technologies accumulate vast amounts of data that exceed meticulous human inspection capabilities. Despite the increasing demands, the full application of machine learning has been hindered by the need for data-specific optimizations. In this study, we introduce a new deep-learning-based phase retrieval method for imperfect diffraction data. This method provides robust phase retrieval for simulated data and performs well on weak-signal single-pulse diffraction data from X-ray free-electron lasers. Moreover, the method significantly reduces data processing time, facilitating real-time image reconstructions that are crucial for high-repetition-rate data acquisition. Thus, this approach offers a reliable solution to the phase problem and is expected to be widely adopted across various research areas.

# 1. Introduction

The phase problem, a well-known inverse problem involving the extraction of phase information hidden in the interference fringes of measured intensities, is prevalent in nature and complicates the direct interpretation of diffraction signals related to target objects[1]. Its impact spans various research modalities, including X-ray crystallography and high-resolution imaging, driving active development to recover lost phase information. While several methods exhibiting good performance have been developed, they often require significant data handling time. The effectiveness of these techniques depends on the completeness of the measured data, noise contamination, and technique-specific challenges in data collection. Deep learning (DL) has shown considerable potential in addressing these issues[2,3]. It reduces processing time by replacing conventional approaches with non-iterative operations after appropriate training, which can be accelerated with graphic processing units (GPUs). Due to these advantages, substantial efforts have been directed toward using DL for denoising, classification, and phase retrieval[4–12], though these tasks remain challenging in X-ray diffraction.

Recent advancements in the development of brighter X-ray sources that provide ultrashort X-ray pulses, such as X-ray free-electron lasers (XFELs), have significantly enhanced the ability to observe ultrafast molecular bonding processes, transient material dynamics, and hidden material phases in strongly driven nonequilibrium states[13–16]. Diffraction imaging, which retrieves phase information through numerical iterations, holds great promise for determining the structure of single specimens. However, the diffraction signals, often plagued by low signal-to-noise ratios due to limited photons and data imperfections, have constrained the practical application of DL for interpreting experimental data[17].

In this study, we propose a new deep neural network (DNN) for phase retrieval of imperfect single-pulse diffraction patterns, enabling real-time image reconstruction for single-particle imaging experiments using XFELs. The network is based on a residual neural network (ResNet) with weight-corrected convolution layers designed to handle diffraction signals[18]. It was trained using masked diffraction patterns as inputs, which were generated from pseudo-random objects without any physical bias. We demonstrated the network's excellent performance with simulated data by comparing it to conventional iterative phase retrieval algorithms. After verifying its effectiveness, we applied the network to single-pulse diffraction data obtained with XFELs, where it exhibited robust real-time image reconstruction with improved image quality. By providing a solution to the phase problem in X-ray diffraction imaging, this study addresses a significant bottleneck in data processing time by eliminating the need for computationally expensive iterative phase retrievals. With advancements in the development of new light sources that offer high brilliance and repetition rates, data accumulation rates have increased exponentially, necessitating rapid data processing. We believe that the proposed DNN method will serve as a crucial basis for advancing scientific discoveries through effective data mining.

2. Results

*2.1. DNN for the phase problem of diffraction patterns*

By replacing computer vision tasks with inference using pretrained parameters, DNNs deliver superior performance on such tasks and have become increasingly efficient with a combination of convolution operations[19]. Various convolution types have been developed to enhance performance for specific applications. Depth-wise separable convolution (DSC) is an efficient convolution method, typically using ten times fewer parameters than plain convolution[20]. Partial convolution (PC) functions as a mask-aware convolution, allocating occluded data based on known data[21]. Fast Fourier convolution (FFC) provides a global receptive field through an additional convolution of the Fourier transform of the input[22]. Recently, a ResNet-based DNN named LaMa has demonstrated exceptional performance in image inpainting, even with large masks[23]. LaMa's straightforward architecture, which includes downscaling layers, residual blocks, and upscaling layers, employs FFC across the entire network to leverage the global receptive field. Building on this basic architecture, we introduce deep phase retrieval (DPR), a new DNN featuring an encoder–decoder architecture with two novel operations: an encoder with weighted partial convolutions (WPC) and a two-stage decoder with intermediate Fourier modulation (Fig. 1b and Supplementary Table 1). DPR is a promising network for the immediate reconstruction of imperfect, noisy diffraction patterns, utilizing WPC and FFC to reflect the nature of X-ray diffraction.

While PC equally distributes known information to missing values within convolving regions, WPC employs a physics-based approach to assign position-dependent weights based on the Guinier–Porod model. This model describes the radial intensity distribution in a small-angle region (see Methods)[24]. As the diffraction intensity typically decreases by $Q^{-4}$ for momentum transfer $\boldsymbol{Q}\ (=\boldsymbol{k}_f - \boldsymbol{k}_i)$ with wave vectors $\boldsymbol{k}_{(i,f)}$ of incoming ($i$) and outgoing ($f$) light, the WPC assigns $Q$-dependent weight factors from the Guinier–Porod model for a smooth sphere to known values during the operation of the PC. In the two-stage decoder, the diffraction-compensated decoder, which performs Fourier modulation before the unit blocks of the residual structure, is connected serially to the base decoder. The Fourier modulation is achieved by replacing the Fourier magnitudes of the primary outputs with the inputs. This operation retrieves the initial diffraction patterns that attenuate through the deep layers of the network, aiding in the accurate generation of Fourier transform pairs. Additionally, FFC operates analogously with conventional phase retrieval algorithms, which iteratively connect Fourier space information using discrete Fourier transform (DFT) and inverse DFT between diffractions and objects. These components work together to reconstruct the lost phase information from imperfect, photon-limited diffraction data.

To generate the dataset, we constructed a diffraction model using pseudo-random objects created from combinations of two pre-existing datasets that separately assigned the shape and density of the objects (see Methods). Diffraction patterns were generated by performing a fast Fourier transform (FFT) along with the diffraction model, incorporating additional operations to account for experimental conditions such as spatial coherence, limited photon counts, and measurement noise. The resulting patterns were then partially obscured with irregular masks, simulating data loss due to pixel arrangements on a detector and a central beam stop that blocks intense direct beams in actual measurements. These patterns were fed into the network as inputs (Fig. 1a). We used the AdamW optimizer for backpropagation with a custom-designed loss function (see Methods)[25].

*2.2. DPR in phase retrievals and evaluation of its performance*

We first validated the improved performance of the WPC-based encoder and diffraction-compensated decoder within the DPR framework. Compared to encoders based on PC and FFC, the WPC-based encoder exhibited lower validation loss during training and showed significant improvements in the *R*-factor ($R_F$), while maintaining or even surpassing the peak signal-to-noise ratio (PSNR) and structural similarity index measure (SSIM) (see Methods, Fig. 1c, d). Additionally, the two-stage decoder including the diffraction-compensated decoder outperformed a single base decoder with doubled residual blocks, despite having 43% fewer total trainable parameters (Fig. 1c, d). This confirms that DPR provides enhanced performance with an efficient architecture and effective handling of imperfect diffraction signals.

We also examined how the reconstruction performance of DPR depends on the size of the training dataset by increasing the size from 12,000 to 2, 4, 8, and 16 times (Fig. 2a). Although a slight improvement was observed in $R_F$, PSNR, and SSIM, the network performance quickly saturated with larger training datasets. This indicates that the current amount of training data is sufficient for DPR, given the efficiency of network training. After validating the DPR architecture, we compared its phase retrieval performance with those of conventional iterative projection algorithms, specifically hybrid input–output (HIO) and generalized proximal smoothing (GPS) (see Methods)[26,27]. We simulated diffraction patterns using the same diffraction model as for the training datasets, applying irregular masks. Additionally, we generated a second set of test data with a constant central mask covering 16 × 16 pixels at the center to compare scenarios with and without substantial data loss. We compared the image reconstructions from DPR; DPR with refinement; and the two conventional phase retrieval algorithms, HIO and GPS. Refinement for DPR involved a few iterations of GPS to obtain the final images (see Methods). The results demonstrate that DPR significantly outperformed HIO and GPS in reconstructing real-space images (Fig. 2b). While HIO and GPS produced indistinct images, especially with large masks, DPR consistently provided superior performance regardless of masking areas. Moreover, DPR reconstructed more detailed features with refinement, though this came at the cost of increased noise from the diffraction signals.

Overall, DPR achieved a higher $R_F$ than GPS for irregular masks but demonstrated significantly better image quality with higher PSNR and SSIM, indicating its robust performance on noisy diffraction signals (Fig. 2c). Additionally, DPR exhibited similarly high levels of PSNR and SSIM in both the irregular and constant center mask cases, highlighting its superior performance in scenarios involving partial data loss. DPR with refinement achieved a significantly lower $R_F$, even lower than that of GPS, though this came at the expense of image quality. For further application to experimental data, the AdamWR optimizer with adaptive sharpness-aware minimization (ASAM) was employed to mitigate failures by improving generalization (see Methods)[25,28]. This approach also led to a notable improvement in the quality of the reconstructed images from the test data (Fig. 2c).

DPR also benefits from having fewer training parameters ($1.52 \times 10^7$ in total) than conventional deep convolutional neural networks, despite the large input size of 512 × 512. It efficiently addresses complex phase problems using WPC and FFC, handles imperfect diffraction signals with an appropriate physical model, and utilizes Fourier-space information similar to conventional phase retrieval methods. The processing times were 9.02 ± 0.00215 ms and 52.2 ± 2.45 ms per data for DPR and DPR with refinement, respectively, using a single NVIDIA GeForce RTX 3090 GPU (Fig. 2d). This represents more than a 1,000-fold increase in speed over that of conventional iterative phase retrieval algorithms, underscoring the superiority of DPR for real-time processing of data from upcoming MHz-repetition-rate XFELs[29].

*2.3. DPR in phase retrievals of experimental data from the XFEL*

After demonstrating the performance with simulated data, DPR was applied to experimental data obtained from XFELs. Single-pulse X-ray diffraction imaging experiments were conducted at the Pohang Accelerator Laboratory-XFEL (PAL-XFEL) (see Methods). In these experiments, X-rays with a photon energy of 5 keV were used, and single-pulse diffraction patterns were recorded by a charge-coupled device (CCD) detector positioned 1.60 m downstream of the sample, providing a pixel resolution of 10.3 nm for a 512 × 512 window. Specimens of Ag nanoparticles, with characteristic flower and cube shapes, were randomly dispersed and mounted on thin $Si_3N_4$ membranes for measurement.

Real-space images were directly obtained from the diffraction signals using DPR (Fig. 3a). The central regions of the diffraction patterns were obscured by a beam stop and strong parasitic scattering near the direct beam. The fringe oscillations from the specimen appeared blurry due to the experimental conditions, including imperfect spatial coherence and other signal contaminations. Despite these challenges, DPR successfully extracted accurate images from the measured diffraction patterns. The images obtained with DPR and DPR with refinement showed distinct shapes with relatively high contrast compared to results from conventional iterative algorithms, HIO and GPS. Notably, DPR is not biased toward low-$Q$ signals near the diffraction center, which represent a significant portion of the total diffraction intensities, resulting in a lower local $R_F$ for high-$Q$ signals (see Methods, Fig. 3a, and Supplementary Fig. 2a). Since the diffraction signals in the high-$Q$ region provide high-resolution information on internal structures, DPR produces real-space images with clearer shapes and more detailed structures.

The results showed strong positive correlations (above 0.8) to those from conventional algorithms, indicating a high degree of agreement in their morphologies (Fig. 3b). An important advantage of the DNN method is that DPR does not require support constraints. It directly converts Fourier-space data into corresponding real-space objects without additional information, unlike conventional phase-retrieval algorithms that require support estimation for real-space constraints. As a hybrid option, refined DPR with 50 iterations of GPS after DPR achieved an improved $R_F$, even lower than the GPS result with a thousand iterations (Supplementary Fig. 2a). This indicates that DPR provides an efficient approach to optimization by giving starting points already close to the global minima.

To further evaluate the phase retrieval performance of DPR on general single-pulse diffraction data, we applied it to public datasets from the Coherent X-ray Imaging Data Bank (CXIDB)[30]. We obtained three datasets, i.e., chlorovirus PBCV-1, bacteriophage T4, and $Fe_2O_3$ ellipsoid nanoparticles, from the repository[31]. In these experiments, X-rays with a photon energy of 1.2 keV were used, and diffraction signals were measured with a 1-megapixel pnCCD positioned 0.738 m downstream of the sample. This setup provided ideal pixel resolutions of 19.9 nm with a 512 × 512 window and 9.93 nm after 2 × 2 binning for the $Fe_2O_3$ ellipsoid dataset.

Despite the completely different samples and experimental conditions, real-space images were successfully obtained using DPR (Fig. 4a). The images produced by DPR exhibited distinct shapes with clear internal structures and higher contrasts compared to those from conventional algorithms. DPR generated real-space objects that were better aligned with high-$Q$ diffraction signals and demonstrated strong positive correlations with the results from conventional algorithms, similar to findings from independent experiments (Fig. 4a, b, and Supplementary Fig. 2b). Thus, DPR was validated as effective for extracting real-space information from diffraction patterns, showing robustness to experimental noise and partial data loss. Notably, DPR was trained without any physical bias and did not require fine-tuning procedures for different types of samples. The consistently improved performance across various datasets confirms the general applicability of DPR. This method enables rapid reconstruction of real-space images from imperfect, noisy diffraction patterns within 10 ms using a single GPU, regardless of experimental conditions or sample types, achieving real-time phase retrieval for XFEL data. Moreover, the techniques employed in DPR, such as WPC, are not limited to phase retrieval but are applicable to solving various problems in X-ray diffraction experiments, including classification and denoising of measured data.

## 3. Discussion

The DNN with the newly proposed architecture excels in solving the phase problem, demonstrating outstanding performance in the phase retrieval of X-ray diffraction patterns. Notably, this network shows excellent tolerance to experimental noise and partial data loss. When applied to single-pulse XFEL diffraction patterns, it achieves rapid and direct reconstruction of real-space images, enabling real-time phase retrieval. The increasing importance of high-speed data processing arises from the large volumes of data generated in a short time by next-generation X-ray sources, and DPR is well-suited to handle such extensive datasets.

The WPC, which utilizes the Guinier–Porod model to guide lost information, highlights the importance of properly handling diffraction data to extract structural information. Despite the WPC-based encoder comprising only 10% of the total trainable parameters in DPR, this approach can be easily adapted to various types of incomplete experimental data, such as X-ray absorption or emission data, by applying appropriate physical models for further improvements in DL-based operations. Thus, DPR not only provides real-time phase retrieval for imperfect diffraction patterns but also represents a novel method for managing partially damaged data from various experiments with distinctive characteristics. The approach is particularly relevant for time-resolved diffraction imaging with high-repetition-rate XFELs, allowing observation of femtosecond dynamics in systems driven far from equilibrium, thus revealing hidden material phases not accessible through equilibrium thermodynamics. DPR is poised to significantly advance this research area by fully utilizing massive datasets in parallel with data collection.

## 4. Methods

### 4.1. Weighted partial convolution

Building on the concept of PC, WPC incorporates position-dependent weights based on the Guinier–Porod model, which describes the radial intensity distribution in small-angle scattering data[21,24]. For an ideal sphere with a smooth surface, the Guinier–Porod model provides the relationship between intensity $I$ and momentum transfer $Q$ as follows:

$$I(Q) = \begin{cases} G \exp\left(-\dfrac{R^2 Q^2}{5}\right), & \text{for } Q \leq Q_1 \\ G \exp\left(-\dfrac{R^2 Q_1^2}{5}\right)\left(\dfrac{Q_1}{Q}\right)^4, & \text{for } Q > Q_1 \end{cases} \qquad (1)$$

where $G$ is the Guinier scale factor; $R$ is the sphere radius; and $Q_1$ is the boundary momentum transfer between the Guinier and Porod models, defined as $Q_1 = \sqrt{10}/R$. Here, $R$ is given by $\pi/\sigma$, where $\sigma$ is the oversampling ratio along an axis, to match a unit of momentum transfer with a pixel of the measured diffraction pattern. The position-dependent weights of the WPC were determined using Eq. (1), with $\sigma = \min(H, W)/64$, where $H$ and $W$ are the height and width of the input for each layer, respectively, and 64 represents the matrix size allocated for the final real-space images. The operation of WPC with convolution kernel $\boldsymbol{K}$ is defined as

$$x' = \begin{cases} \boldsymbol{K}^T(\boldsymbol{X} \odot \boldsymbol{M})\dfrac{\sum_i \boldsymbol{W}_i}{\sum_{M_i \neq 0} \boldsymbol{W}_i}, & \text{if } \sum_i \boldsymbol{M}_i \neq 0 \\ 0, & \text{otherwise} \end{cases} \qquad (2)$$

where $\boldsymbol{X}$ is the input, $\boldsymbol{M}$ is the binary mask for the valid data points, and $\boldsymbol{W}$ is the weight in the region covered by the kernel during convolution.

*4.2. Diffraction model*

The diffraction model generates diffraction patterns from objects, reflecting the properties of single-pulse X-ray diffraction imaging experiments using XFELs. Basic diffraction patterns are produced by taking the absolute square of the FFT of pseudorandom objects derived from a combination of preexisting datasets: EMNIST and CIFAR-100[32,33]. EMNIST consists of handwritten characters that define the shapes of the objects, while CIFAR-100 includes images from 100 classes that provide internal density distributions. Specifically, EMNIST images are enlarged using maximum filters with random widths ranging from 3 to 7 pixels, then modified by affine transforms with random angles (0° to 90°) and scales (0.8 to 1.5), and finally cropped to 64 × 64 pixels. This results in an oversampling ratio of approximately 10 to 20 within a 512 × 512 window. CIFAR-100 images are cropped with random scales and aspect ratios ranging from 0.08 to 1 and 0.75 to 1.33, respectively, and then resized to 64 × 64 pixels. After generating the basic diffraction patterns, the Gaussian Schell model is used to account for the finite spatial coherence length of the radiation from the XFELs, as follows:

$$I' = \left|\mathrm{FT}\left\{\mathrm{FT}^{-1}[I] \odot \exp\left(-\frac{r^2}{4\sigma_\mu^2}\right)\right\}\right| \qquad (3)$$

where $I$ is the diffraction pattern, $r$ is the matrix of radial distances from the center, and $\sigma_\mu$ is the spatial coherence length[34]. $\sigma_\mu$ is given by 200 pixels with 10% random deviations. Then, the diffraction patterns are scaled to have total diffraction intensities in the range of $10^6$–$10^7$, and mixed Poisson–Gaussian noise was added to the patterns as follows:

$$I'_i = \mathrm{Pois}\left(I_i \cdot \frac{I_{\mathrm{total}}}{\sum_j I_j}\right) + \mathcal{N}(0, \sigma) \qquad (4)$$

where Pois(λ) generates random values from a Poisson distribution with λ events, and $\mathcal{N}(\mu, \sigma)$ generates random values from a normal distribution with mean $\mu$ and standard deviation σ. The final diffraction patterns were paired with random masks. These masks were created using a combination of center masks with random radii (ranging from 8 to 32 pixels) and positional deviations (ranging from −8 to 8 pixels along each axis), along with irregular masks from the NVIDIA Irregular Mask Dataset[21]. The occlusion ratio for the irregular masks was limited to 50%. The total number of generated patterns was 96,000 for training, 12,000 for validation, and 12,000 for testing.

### 4.3. Loss function and network training

The loss function comprises the mean absolute error (MAE), MAE of the gradient, perceptual loss, and $R_F$ of the ground-truth Fourier magnitudes. These functions are defined as

$$\mathcal{L}(X,Y) = \frac{1}{N}\sum_{i=1}^{N}|X_i - Y_i|, \quad \mathcal{L}_{\text{grad}}(X,Y) = \frac{\sum_{Y_i \neq 0}\|\nabla X_i - \nabla Y_i\|_1}{\sum_{Y_i \neq 0} 1}$$
$$\mathcal{L}_{\text{perc}}(X,Y) = \mathcal{L}(\Phi[X], \Phi[Y]), \quad R_F^{\text{GT}}(X,Y) = \frac{\sum_i ||\text{FT}[X]|_i - |\text{FT}[Y]|_i|}{\sum_i |\text{FT}[Y]|_i}$$
(5)

where $X$ is the output from the network, $Y$ is the target, and $\Phi$ is the pretrained neural network. For perceptual loss, intermediate outputs after the 4th and 5th blocks of ImageNet-pretrained VGG-19 were used[35]. Additional weights were applied to the outputs based on the square root of the total diffraction intensity to reduce the influence of weak data. After an ablation study of each component, the final loss function was defined as $\mathcal{L}_{\text{total}} = \mathcal{L} + 10\mathcal{L}_{\text{grad}} + 0.1\mathcal{L}_{\text{perc}} + 0.01R_F^{\text{GT}}$ (Supplementary Fig. 1). Based on the loss function, the network is trained by the AdamW optimizer with $\beta_1 = 0.9$, $\beta_2 = 0.999$, and a weight decay of 0.0001 for 500 epochs followed by 100 epochs with learning rates of 0.001 and 0.0001, respectively[25]. For the case using the AdamWR optimizer with ASAM, parameters of ASAM were set as $\rho = 0.2$ and $\eta = 0.01$; the learning rate was determined by cosine annealing with a warm restart scheduler as $\alpha_i = \alpha_{\min} + 0.5(\alpha_{\max} - \alpha_{\min})(1 + \cos(\pi T/T_i))$, where $\alpha_{\min} = 10^{-8}$, $\alpha_{\max} = 0.005$, $T$ is the number of epochs after a recent restart, and $T_i$ is the number of epochs between two restarts, initially set to 40 and doubled after each restart[28]. Twelve NVIDIA GeForce RTX 3090 GPUs were used for network training.

### 4.4. Evaluation metrics

The performance of DPR was evaluated using three metrics: $R_F$, PSNR, and SSIM[36]. The metrics are defined as follows:

$$R_F(X,I) = \frac{\sum_{i,\text{valid}}||\text{FT}[X]|_i - \sqrt{I_i}|}{\sum_{i,\text{valid}}\sqrt{I_i}}, \quad \text{PSNR}(X,Y) = 20\log_{10}\frac{\max(Y)}{\sqrt{\frac{1}{N}\sum_{i=1}^{N}(X_i - Y_i)^2}}$$
$$\text{SSIM}(X,Y) = \frac{(2\mu_X\mu_Y + c_1)(2\sigma_{XY} + c_2)}{(\mu_X^2 + \mu_Y^2 + c_1)(\sigma_X^2 + \sigma_Y^2 + c_2)}$$
(6)

where $X$ is the output from the network, $Y$ is the target, $I$ is the diffraction pattern, $\mu_X$ is the mean of $X$, $\sigma_X^2$ is the variance of $X$, and $\sigma_{XY}$ is the covariance of $X$ and $Y$. $c_1$ and $c_2$ in SSIM are given by $(0.01\max(Y))^2$ and $(0.03\max(Y))^2$, respectively. A two-sided Mann–Whitney U test was also performed using the evaluation metrics to identify statistical differences in DPR. For cases involving experimental data, the local $R_F$ and Pearson correlation coefficients (PCCs) for all pairs were calculated. The local $R_F$ was calculated pixelwise for data points with photon counts exceeding 0.5, while the PCC was defined as $\mathrm{PCC}(X,Y) = \sigma_{XY}/\sigma_X\sigma_Y$.

*4.5. Phase retrieval parameters*

For phase retrieval using HIO and GPS, 1000 iterations were performed with 100 initial random phases[26,27]. The HIO algorithm was employed with $\beta = 0.9$, and the error reduction algorithm accounted for 10% of the total iterations. GPS was executed as R variants (GPS-R) with the following parameters: $t = 1$, $s = 0.9$, $\sigma$ linearly increasing from 0.01 by a factor of 10 at 40% and 70% of the total iterations, and $\gamma = 1/2\alpha^2$ with $\alpha$ linearly decreasing from 1024 by 10% every 100 iterations. Both algorithms also used the shrink-wrap algorithm with $\sigma$ linearly decreasing from 3 pixels by 1% and a threshold of 20% of the maximum value to update the support constraints every 50 iterations. The initial supports were 60 × 60 pixels for the test data and 30 × 30 pixels for the experimental data. The final images were selected based on $R_F$: a single image for the test data and an average of five images for the experimental data. To refine the outputs from DPR, the support constraints were derived from the output images by thresholding at 1% of the 99th percentile values. Using these supports, GPS-R was conducted for 50 iterations with the following parameters: $t = 1$, $s = 0.9$, $\sigma$ increasing from 0.1 to 1 at 40% of the total iterations, and $\gamma = 1/2\alpha^2$ with $\alpha$ linearly decreasing from 1024 by 20% every 10 iterations.

*4.6. Single-pulse X-ray diffraction imaging experiments*

The experiments were conducted at the nanocrystallography and coherent imaging (NCI) beamline of the PAL-XFEL[37]. X-ray pulses from self-amplified spontaneous emission with a nominal photon energy of 5 keV and a bandwidth of $\Delta E/E \approx 5 \times 10^{-3}$ were used for the experiments. The X-ray pulses were focused into a 5 μm (horizontal) × 7 μm (vertical) area by a pair of Kirkpatrick–Baez mirrors installed 5 m upstream of sample position, giving an effective photon flux of approximately $8 \times 10^9$ photons·μm$^{-2}$ per pulse. Diffraction patterns were recorded using a 1-megapixel multi-port CCD with a pixel size of 50 × 50 μm$^2$, located 1.6 m downstream of the sample position. A beam stop was placed in front of the detector to block the direct X-ray beam and cover a quadrant of the detector plane. The samples included Ag flower and cube nanoparticles with approximate widths of 150 nm and 100 nm, respectively. These were spread on 100-nm-thick $Si_3N_4$ membranes and loaded into the imaging chamber. All beam paths, including the imaging chamber, were kept under vacuum during the measurements. Background signals were subtracted from the measured diffraction patterns, and missing values were substituted with values at centrosymmetric positions in accordance with Friedel's law.

# Figures

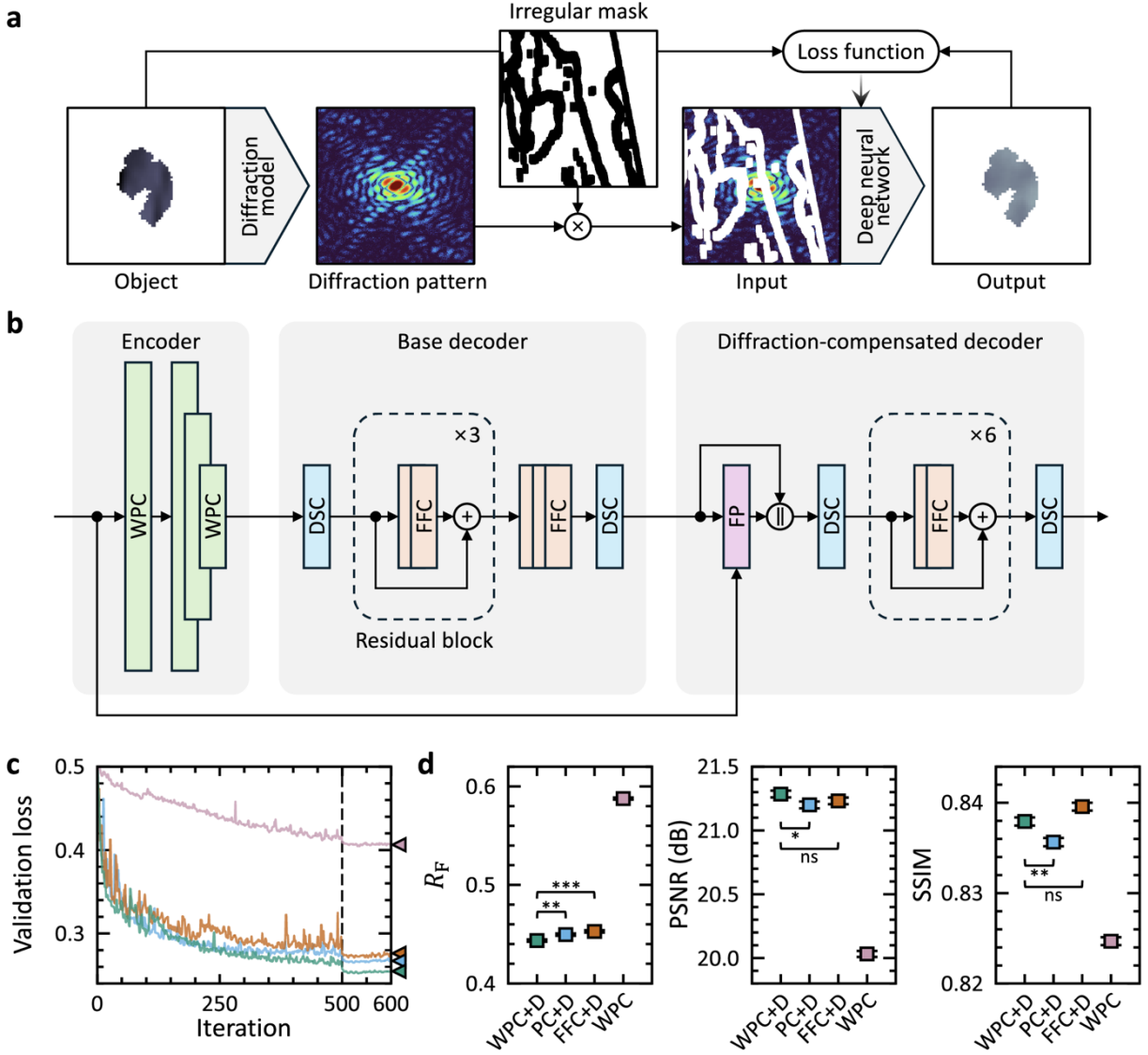

**Fig. 1. DNN for real-time phase retrieval of imperfect single-pulse diffraction patterns. a**, Schematic diagram of data generation and the network training. **b**, Schematic diagram of DPR. The network consists of a WPC-based encoder and a two-stage decoder, including a base decoder and a diffraction-compensated decoder (+D). **c,d**, Evolution of validation loss during training iterations (**c**) and evaluation metrics (**d**) for WPC-, PC-, and FFC-based encoders with and without +D. The boxes and whiskers represent the average and standard errors of each metric, respectively. Differences with WPC+D are verified by the Mann-Whitney U test (not indicated, $p \leq 10^{-8}$; ***$10^{-8} < p \leq 0.001$; **$0.001 < p \leq 0.01$; *$0.01 < p \leq 0.05$; ns, $p > 0.05$).

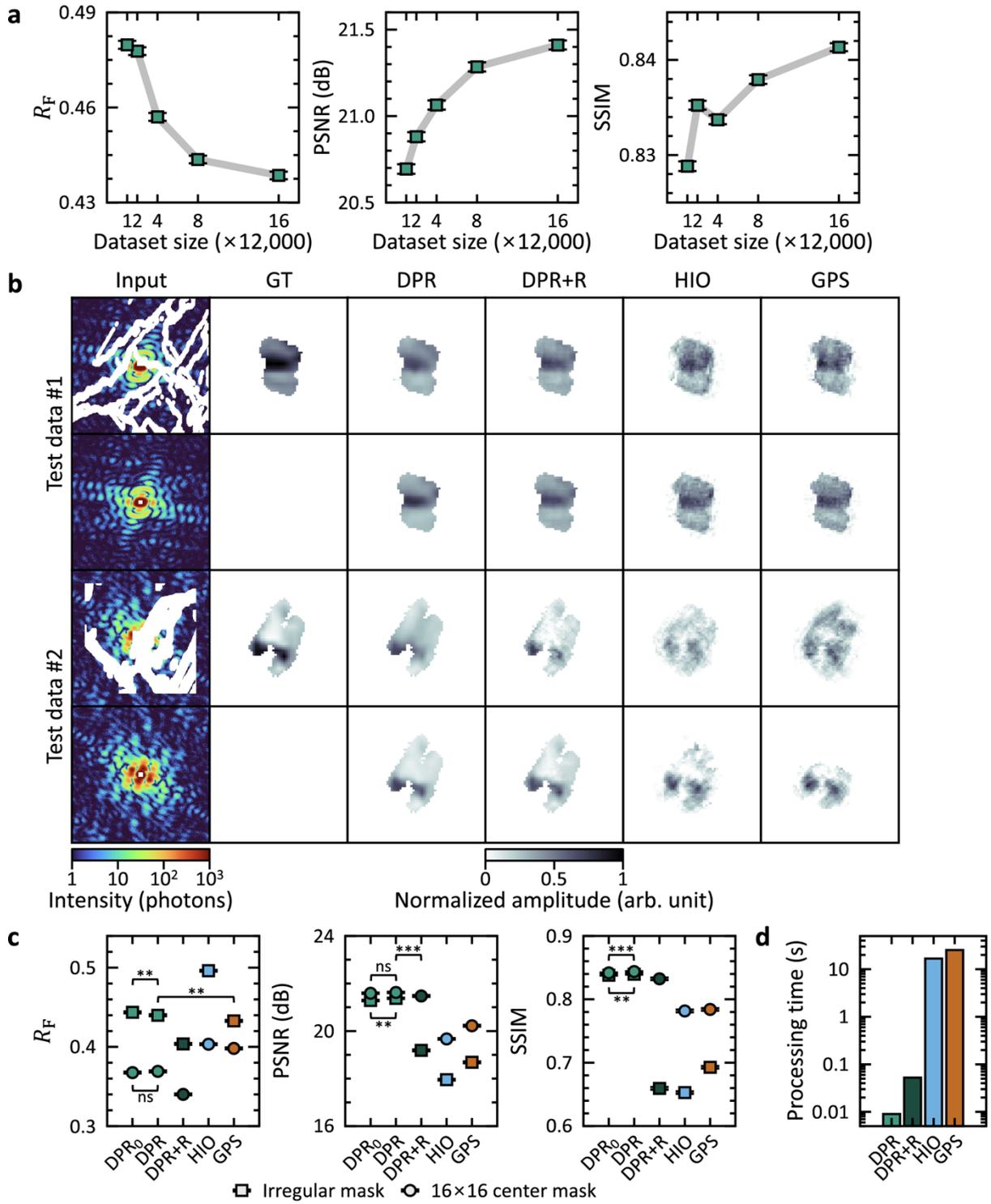

**Fig. 2. Phase retrieval performance of DPR on simulated data. a**, Change in evaluation metrics for different sizes of the training dataset. **b**, Examples of real-space image reconstructions from simulated data. Diffraction patterns were simulated using the diffraction model with irregular masks and a constant central mask. Reconstructed real-space images from each algorithm are compared with ground-truth (GT) images. **c,d**, Comparisons of evaluation

metrics (**c**) and processing times (**d**) between HIO and GPS, as representative conventional algorithms, DPR trained with the AdamW optimizer (DPR$_0$), DPR trained with the AdamWR optimizer and ASAM (DPR), and DPR with refinement (DPR+R). The boxes and whiskers represent the average and standard errors of each metric, respectively. Differences with DPR are verified by the Mann-Whitney U test (not indicated, $p \leq 10^{-8}$; ***$10^{-8} < p \leq 0.001$; **$0.001 < p \leq 0.01$; ns, $p > 0.05$).

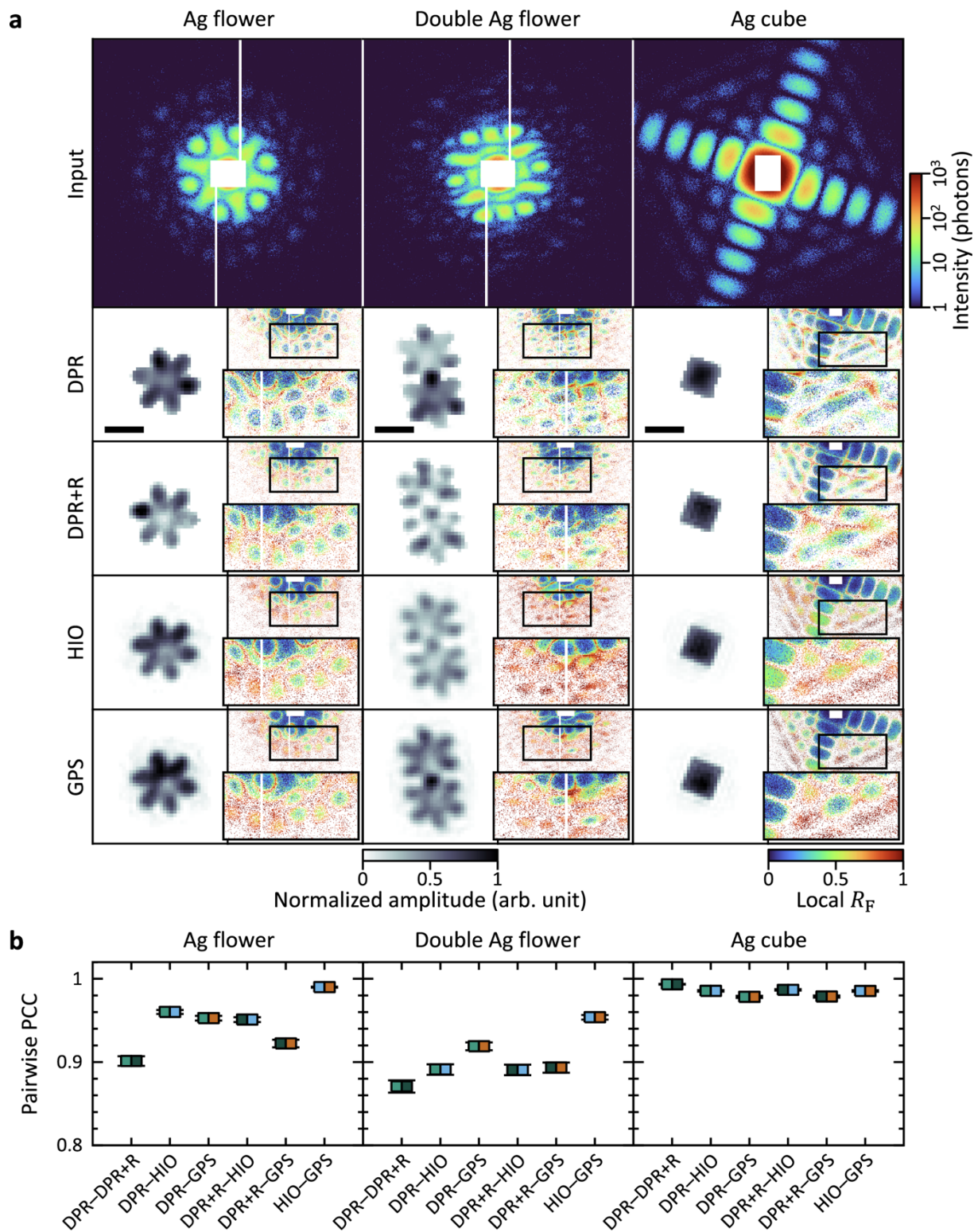

**Fig. 3. DPR on data from single-pulse X-ray diffraction imaging experiments using the XFEL. a**, Comparisons of reconstruction results from diffraction patterns of Ag flower, double Ag flower, and Ag cube nanoparticles measured at PAL-XFEL. The first row displays the measured single-pulse diffraction patterns, with the central part blocked by the beam stop and missing data along vertical line gaps due to detector chip arrangements. Phase retrievals using DPR, DPR with refinement (DPR+R), HIO, and GPS were performed, and the reconstructed images are shown in the left column beneath each diffraction pattern. Local $R_F$ distributions, derived from the Fourier magnitudes of the images, are presented in color in the right columns. Scale bars represent 100 nm. **b**, Pairwise Pearson correlation coefficients (PCCs) for all pairs of results obtained using each method. The boxes and whiskers represent PCCs and their confidence intervals at a 95% confidence level.

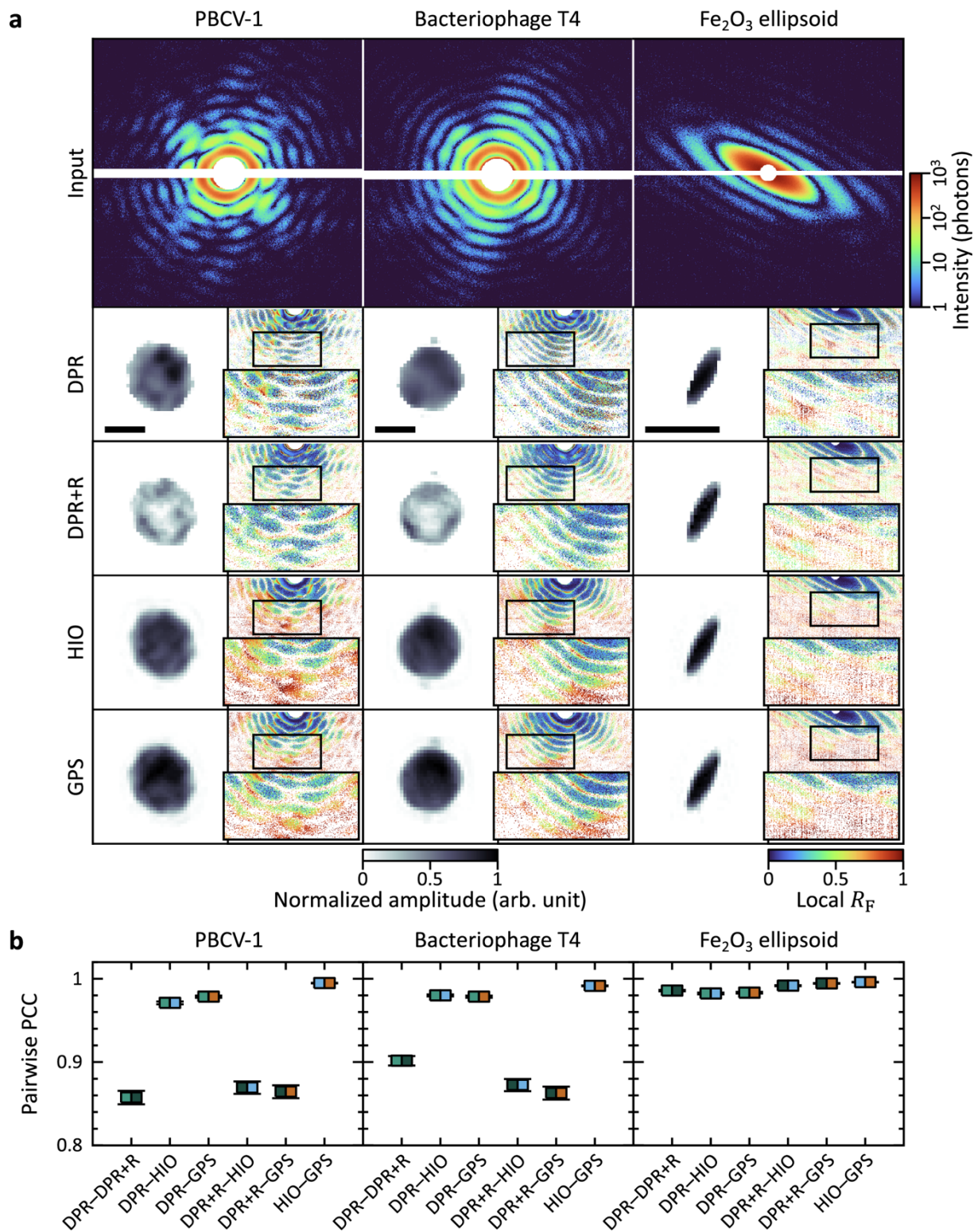

**Fig. 4. DPR on publicly available single-pulse X-ray diffraction data. a**, Comparisons of reconstruction results from diffraction patterns of chlorovirus PBCV-1, bacteriophage T4, and Fe$_2$O$_3$ ellipsoid nanoparticles from CXIDB. Measured diffraction patterns served as inputs, with masked regions not recorded due to detector chip arrangements. For Fe$_2$O$_3$ ellipsoid

nanoparticles, 2 × 2 binning was applied to reduce the oversampling ratio before phase retrieval. Phase retrievals using DPR, DPR with refinement (DPR+R), HIO, and GPS were compared. Reconstructed images for each phase retrieval method are displayed in the left column beneath the diffraction patterns in the first row. Local $R_F$ distributions, derived from the Fourier magnitudes of the images, are shown in color in the right columns. Scale bars represent 200 nm. **b**, Pairwise PCCs for all pairs of results obtained using each method. The boxes and whiskers represent PCCs and their confidence intervals at a 95% confidence level.

**Data Availability**

The data supporting the findings of this study are available from the corresponding authors upon a request.

**Code Availability**

The source code and trained parameters of DPR are available at https://github.com/sungyun98/DPR.

**Acknowledgement**

S.Y.L. would like to thank Jaechang Kim from POSTECH for helpful discussion on DL. The experiments at PAL-XFEL were approved by Korean Synchrotron Users Association. This work was supported by the National Research Foundation of Korea (Grant Nos. 2015R1A5A1009962, 2022M3H4A1A04074153, and RS-2024-00346711).

**Author Information**

**Contributions**

C.S. conceived the project. S.Y.L. developed the DNN for phase retrieval of imperfect single-pulse diffraction patterns and conducted all related procedures. D.H.C., C.J., D.S. D.N., S.K., and C.S. performed the single-pulse X-ray diffraction imaging experiments at PAL-XFEL. S.Y.L. and C.S. wrote the manuscript with inputs from all authors.

**Corresponding authors**

Correspondence to Changyong Song.

**Ethics Declarations**

**Competing interests**

The authors declare no competing interests.

**Referenecs**

# Supplementary Information

# Deep-learning real-time phase retrieval of imperfect diffraction patterns from X-ray free-electron lasers

Sung Yun Lee,[1,2,3] Do Hyung Cho,[1,2] Chulho Jung,[1,2] Daeho Sung,[1,2] Daewoong Nam,[4] Sangsoo Kim,[4] Changyong Song[1,2,3,*]

[1]*Department of Physics, POSTECH, Pohang 37673, Republic of Korea, [2]Photon Science Center, POSTECH, Pohang 37673, Republic of Korea, [3]Center for Ultrafast Science on Quantum Matter, Max Planck POSTECH/Korea Research Initiative, Pohang 37673, Republic of Korea, [4]Pohang Accelerator Laboratory, POSTECH, Pohang 37673, Republic of Korea*

\* Corresponding author; email: cysong@postech.ac.kr

| Layer no. | Structure | Output shape (C × H × W) |
|---|---|---|
| 0 | | 1 × 512 × 512 |
| 1 | WPC(7, 1, 3) + BN + ReLU | 64 × 512 × 512 |
| 3 | WPC(3, 2, 1) + BN + ReLU | 128 × 256 × 256 |
| 4 | WPC(3, 2, 1) + BN + ReLU | 256 × 128 × 128 |
| 5 | WPC(3, 2, 1) + BN + ReLU | 512 × 64 × 64 |
| 6 | DSC(3, 1, 1) + BN + ReLU + Split | (256 + 256) × 64 × 64 |
| 7–12 | FFC residual blocks × 3<br>$(x_0, x_0') \rightarrow$ FFC(3, 1, 1) + BN + ReLU $\rightarrow (x_1, x_1')$<br>$(x_1, x_1') \rightarrow$ FFC(3, 1, 1) + BN + ReLU $\rightarrow (x_2, x_2')$<br>$\rightarrow (x_0 + x_2, x_0' + x_2')$ | (256 + 256) × 64 × 64 |
| 13 | FFC(3, 1, 1) + BN + ReLU | (128 + 128) × 64 × 64 |
| 14 | FFC(3, 1, 1) + BN + ReLU | (64 + 64) × 64 × 64 |
| 15 | FFC(3, 1, 1) + BN + ReLU | (32 + 32) × 64 × 64 |
| 16 | Concatenate + DSC(3, 1, 1) + Sigmoid | 1 × 64 × 64 |
| 17 | $x_0 \rightarrow$ Fourier projection $\rightarrow x_1$<br>$(x_0, x_1) \rightarrow$ Concatenate $\rightarrow x_2$ | 2 × 64 × 64 |
| 18 | DSC(3, 1, 1) + BN + ReLU + Split | (32 + 32) × 64 × 64 |
| 19–30 | FFC residual blocks × 6 | (32 + 32) × 64 × 64 |
| 31 | Concatenate + DSC(3, 1, 1) + Sigmoid | 1 × 64 × 64 |
| Total trainable parameters: $1.52 \times 10^7$ | | |

**Supplementary Table 1. Detailed structure of DPR.** Full list of components and output shapes for each layer of the DPR architecture. The layers in encoder, base decoder, and diffraction-compensated decoder are colored in green, orange, and blue, respectively. All convolution operations operate without bias, and their parameters, kernel size, stride, and padding, are noted in parentheses in order. Batch normalization (BN) and rectified linear unit (ReLU) are used in all convolution layers except for the last layers of the decoders, which use sigmoid functions without BN. Split and concatenation are conducted on the channel dimension. The Fourier projection replace the Fourier magnitude of the inputs with square roots of initial diffraction patterns. Final outputs are scaled by initial diffraction patterns.

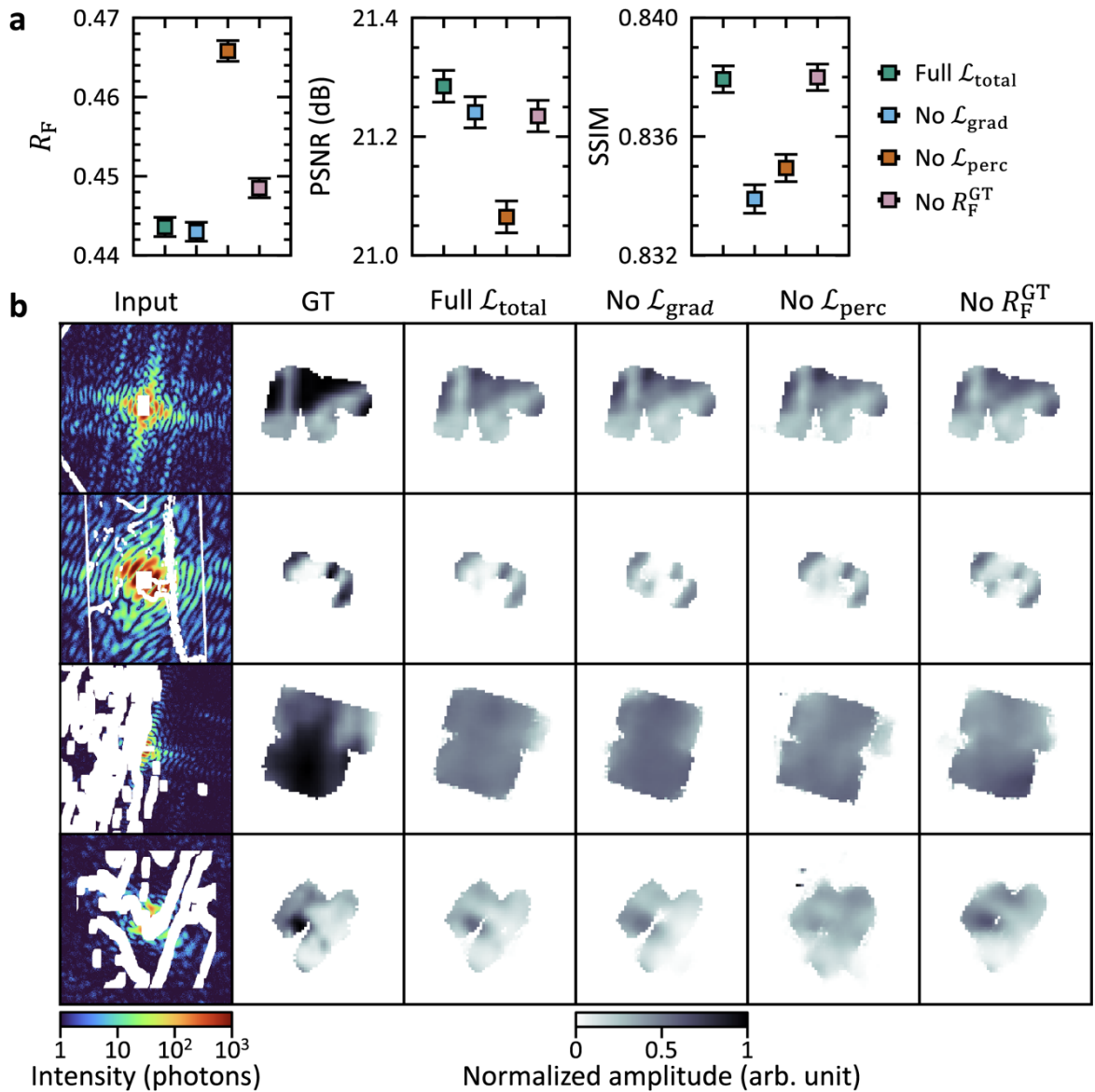

**Supplementary Fig. 1. Ablation study on loss function. a,b**, Comparisons of evaluation metrics (**a**) and example reconstruction results (**b**) between the full loss function and loss functions excluding each component. The boxes and whiskers indicate average and standard error of each metric, respectively. The full loss function offers the best overall performance.

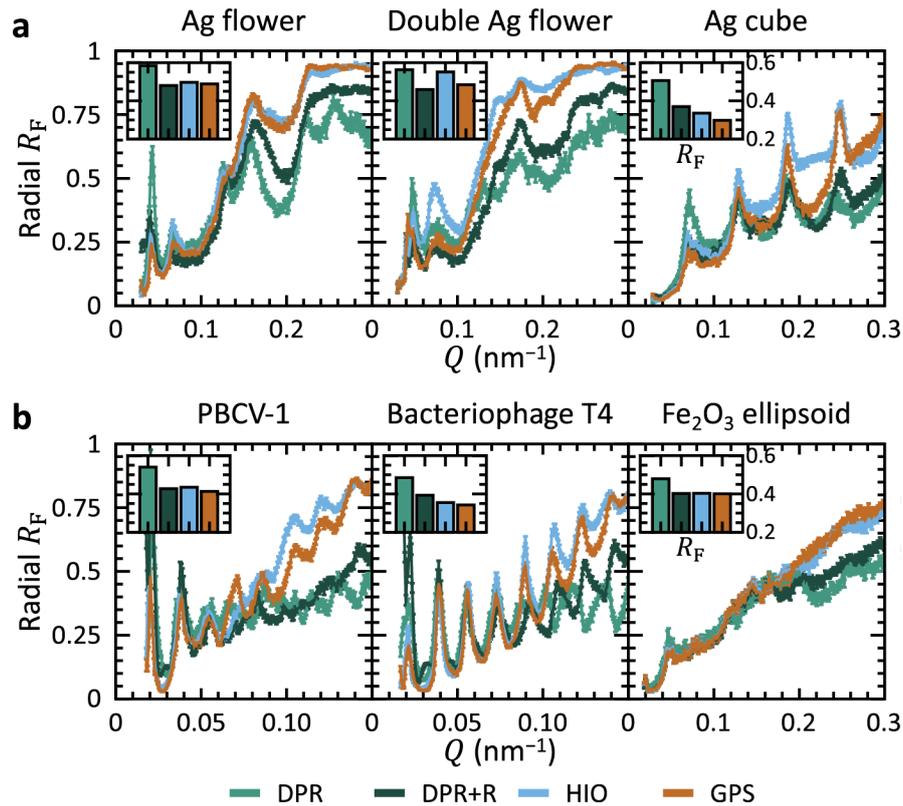

**Supplementary Fig. 2. Local $R_F$ for experimental data. a,b**, Radial distributions of local $R_F$, derived from the reconstructed images of experimental data measured at PAL-XFEL (**a**) and publicly available data deployed at the CXIDB (**b**). Original $R_F$ values are also provided for each method in the inset figures. The boxes and whiskers represent the average and standard errors of each metric, respectively. DPR and DPR with refinement offer a significantly lower level of local $R_F$ for larger momentum transfer $Q$.